# A Recruitment and Human Resource Management Technique Using Blockchain Technology for Industry 4.0

## Md Mehedi Hassan Onik[1,*], Mahdi H. Miraz[2], Chul-Soo Kim[3]

[1,*]Department of Computer Engineering, Inje University, Gimhae-50834, South Korea, Email: hassan@oasis.inje.ac.kr
[2]School of Computer Studies, AMA International University BAHRAIN (AMAIUB), Bahrain, Email: m.miraz@amaiu.edu.bh
[3]Department of Computer Engineering, Inje University, Gimhae-50834, South Korea, Email: charles@inje.ac.kr

## Abstract

Application of Information Technology (IT) in the domain of Human Resource Management (HRM) systems is a sine qua non for any organization for successfully adopting and implementing Fourth Industrial Revolution (Industry 4.0). However, these systems are required to ensure non-biased, efficient, transparent and secure environment. Blockchain, a technology based on distributed digital ledgers, can help facilitate the process of successfully effectuating these specifications. A detailed literature review has been conducted to identify the current status of usage of Information Technology in the domain of Human Resource Management and how Blockchain can help achieve a smart, cost-effective, efficient, transparent and secure factory management system. A Blockchain based Recruitment Management System (BcRMS) as well as Blockchain based Human Resource Management System (BcHRMS) algorithm have been proposed. From the analysis of the results obtained through the case study, it is evident that the proposed system holds definite advantages compared to the existing recruitment systems. Future research directions have also been identified and advocated.

## 1 Introduction

The Blockchain technology is considered as one of the major catalysts transforming the next generation of industry structure. This digital ledger technology is integrating the computational and physical components of the smart industry by providing a fair, cost-effective, accurate, traceable and secure system. However, current industries are facing tremendous pressure to maintain a tenable human resource management system. Since the world is now on its way to adopt another industrial revolution (industry 5.0), industries of all level are in need of a smart employee hiring and management system. Internet-based and computer-aided human resource management system has already been a popular way of hiring and employee evaluation. Gascó [1] in 2004, described, using a case-study conducted at Telefonica which is a renowned telecommunication company in Spain, how the use of Information Technology (IT) can enhance the overall performance of Human Resources Management (HRM) systems.

However, the authenticity of human resource information directly affects the motivation, rate and efficiency of human resource management. Job seekers may hide unpleasant results and present fraudulent information during the recruitment process. Some job applicants submit inflated resume supplementing with fake training and diploma certificates, references, awards, promotions and so forth whereas others deliberately exaggerate their qualifications and abilities [2]. A recent (2017) survey by business review Australia identified that a $250,000 monetary loss to the hiring company can happen yearly due to a bad hiring of an employee on an annual salary $100,000 [3]. In a study by Robert Half [4] conducted amongst Human Resource (HR) Managers in Australia identified that an unsuccessful hiring in an organization may negatively affect productivity (55%), lower staff morale (23%), and cause financial loss (19%). As a result, organizations end up paying a huge amount of cost just to get rid of bad hiring. For example, Amazon pays employees $5,000 bonus for cancelling a job contract [5]. Another survey reveals that 74 percent of employers had wrong hiring. It was also identified that 45 percent worker's skills did not match with their claim and 33 percent had lied about their qualifications [6]. Survey founds, 10 percent out of 2,257 hiring managers stated that they do not have adequate tools to find the right person to hire [7]. Associated Press news stated, India faces a very high rate of employment through fake certificates stating one single state had identified total 1832 such cases in the year 2017 [8].

On the contrary, a trustable, middlemen less, independent and transparent recruitment with efficient HRM systems are a precondition for the successful implementation of Fourth Industrial Revolution (Industry 4.0), or even preparation for the Fifth (Industry 5.0). Blockchain technology possesses significant potentials to eliminate these problems. Blockchain technology has already taken off in several industries because of the fact that it can offer a smart, well- connected, self-organized, transparent, decentralized and immutable system. In this study, we propose a fast, efficient, transparent recruitment and human resource management system using Blockchain (BC) to reduce the risk faced by human resource authority. The proposed system, thereby, provides authentic and effective decision support information for the human resource management of an organization.

## 2 Literature Review

In this Section, we briefly discuss Blockchain, existing recruitment and human resource management approaches and existent issues related to the recruitment process and HRM.

**Blockchain**: Initially, "Bitcoin (BTC)" used to be the only dominant application of Blockchain technology [9]. Yet, with further technical growths, this can be used in other sectors too. Major industries are finance and insurance, health and medical, Internet of Things (IoT), academia, copyright, vehicle sharing, supply chain management, energy management, gambling, government, retail [10-12]. In a Blockchain system, every transaction is stored in a block. Other important parts include: header, transaction counter and transactions or data. Header reserves all information regarding the characteristics and version of that particular block. Transaction counter stores the number of the associated block. Finally, transaction sections store every kind of data related to a particular transaction or action. A Blockchain operation fundamentally consists of four successive steps. Firstly, the sender or initiator announce the transaction publicly. Secondly, members of the sender's Blockchain group validate both the sender and the receiver to approve the transaction. These steps go through several consensus or approval algorithms such as: Proof of Work (PoW), Proof of Stake (PoS), Proof of Activity, Ripple, Practical Byzantine Fault Tolerance (PBFT) etc. [13]. Thirdly, if successfully verified and approved, the transaction takes place between the concerned parties. Finally, after double hashing of data, the block containing latest transaction is added to the existing chain of blocks [13-15].

Yuan [16] presented a Blockchain based Intelligent Transportation Systems (ITS). In this study authors proposed a seven-layer conceptual model for transport management system similar to newly parallel transportation management systems (PTMS). They proposed a ride sharing, logistics and asset management system. Another study proposed a distributed database based energy management system for a direct current (DC) residential distribution system. It proposed a Centralized Energy Management System (CEMS) to control the distribution of electricity in a secure and transparent system. Machine-to-machine (M2M) energy distribution, tracking, validation and billing were performed by an application [17]. A proof-of-concept implementation was demonstrated using a scenario where two electricity producers and one consumer completed their trading amongst each other over a Blockchain ecosystem [18]. Sun [19] proposed a Blockchain based sharing system among human, technology and organization in an IoT enabled smart city.

**HR and Recruitment:** Kavanagh [20] and Lussier [21] in their books discussed not only ongoing human resource management technology but also other technologies like mobile devices and social media driving changes to recruitment and HRM system. They discussed web-based recruitment, HR software, employee evaluation system and relevant issues faced by the current industries. Analoui [22] elaborates the changing patterns of human resource management system. It reinforced the need for a secure and transparent human resource management system. Several issues related to biased hiring and promotion were also highlighted. Already established third-party recruitment systems were also found to have a variety of flaws which made modern business recruitment system more challenging. A study identified 5 key steps where third-party recruitment systems face problems and double up the total cost. These are sequential as follows: cost, process, candidate promotion, contracts and cost again [23]. Marvel [24] demonstrated how technology adoption in HR can reduce the gap between technology and human. Companies end up spending a lot of money a due to inefficient maintenance of network equipment to store and manage HR related information [25].

The major evolutions in HRM systems so far are: addition of mainframe computer in 1970, introduction of ERP system in 1979, advance web-based system in 1990, cloud-based human resource solution in 2006, video hiring system in 2016, and big data analysis in 2017. Current HR and recruiting information handling software are either single server or cloud-based which takes time with security flaws as there as well [26]. Bolman [27] discusses the changing patterns of human resource management system and role of technology for this dynamic change [22]. Major challenges for recruitment, as identified by several studies are: fake information, inefficient promotion and salary management, false reference and break of contract information between employee and employers and finally wrong recruitment [7, 28-29]. Software security flaws and financial data tampering are other key concern; there were even several cases of demanding ransom by hackers from the attacked human resources organizations [30]. Data-stealing and breaching incidents, such as United States (US) Office of Personnel Management (OPM) data breach in June 2015, are alarming and demand a secure recruitment and HR system [31]. Our literature search indicates that Blockchain has not been used for HR before and thus it ensures novelty of our work.

## 3 Proposed Method

In this section, we advocate our proposed methodology for Blockchain based Recruitment Management System (BcRMS) and Blockchain based Human Resource Management System (BcHRMS). Proposed methods are secure, independent, and transparent targeting fourth industrial revolution (industry 4.0) and smart cities.

### 3.1 Proposed Architecture

The following figures (fig.1, fig.2 and fig. 3) pictorially presents the architectural overview of our proposed system. As shown in fig. 1, our proposed system receives the list of applicants from a particular industry. After validation and verification from applicant's previous affiliation and block information, it goes for profiling according to company's requirement. Finally, the system provides a ranked list of verified candidates to the company to make their decision. In short, a company provides a list of applicants and our proposed system provides a list of verified and ranked list of applicants (fig. 1 and fig. 2). During the verification process, our system automatically verifies a profile from the applicant's last workplaces and contracts. If our proposed system (**BcRMS** and **BcHRMS**) finds any irrelevant data, behavioural problems, law-related issues and fake certificates, it will discard those profiles and rank others.

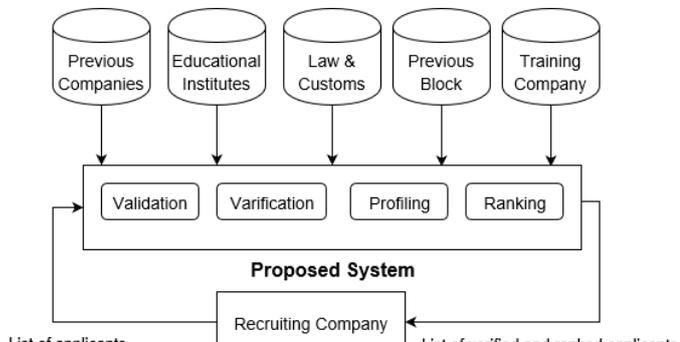

Fig. 1: Blockchain based Recruitment Management System (BcRMS)

As shown in fig. 2, upon successful verification and profiling, the BcRMS system ranks the candidate profiles based on a matching score comparing with the requirements as set by the individual companies in their respective databases. The databases are then updated with the rank scores.

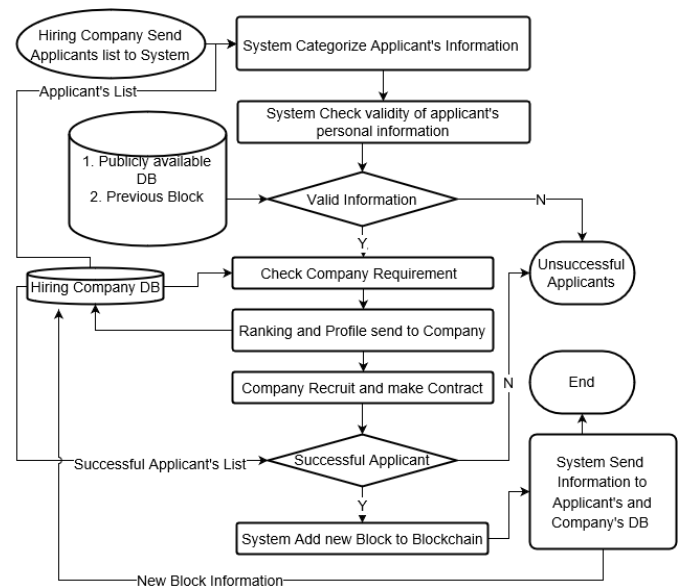

Fig. 2: Blockchain based Recruitment Management System and Blockchain based Human Resource Management System

Finally, when the company receives the ranked and verified applicants list, it makes recruit decisions which are finalized by preparing and signing an employment contract between the employee and the employer. Subsequently, proposed system adds a new block (in this case block 3, as shown in fig. 3 to the existing Blockchain containing the contract information, obviously after achieving consensus from all the concerned parties. For adding a new block, all nodes (employee, organizations, employers etc.) participate in verifying a contract by using the consensus algorithm. This ensures that every new block is well verified and recognized by all.

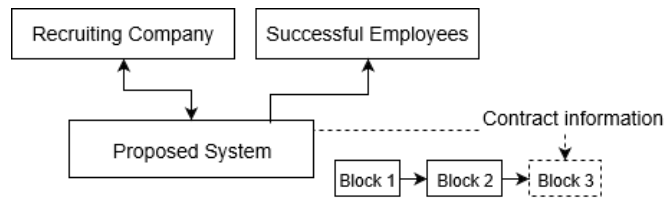

Fig. 3: Blockchain based Human Resource Management System (BcHRMS)

### 3.2 Proposed Algorithm

**Algorithm 1,** implemented in our proposed our hiring system. That has been used to validate and rank the applicant's information or profile. The algorithm has a list of applicants and list of databases to check the information. The authenticity of an applicant's information is verified by using any consensus algorithm discussed in the previous section. Upon successful validation, a ranked list is created based on the company's set requirement.

| Algorithm 1: BcRMS (Blockchain based Recruitment Management System) |
|---|
| **1 Input**: D, where, D = {$D_1, D_2, D_3 \ldots D_n$};   // Database = Companies, Health, Law, Block etc.<br>**2** L, List of Applicants<br>**3 Output**: RL, Ranked List of D     // verified and ranked<br>**4 Begin**<br>**5 for** each $L_i$, i = 1:n do           // check all applicants<br>**6**    **for** each $D_j$, j= 1:m do         // check database<br>**7**       T=Proposed system ($L_i$, $D_j$) // RR Consensus<br>**8**       If (T) add $L_i$ to List RL       // add applicants to list<br>**9**    end<br>**10** end<br>**11** RL = **Sort** (RL)     //Ranked list of Verified applicants<br>**12** End |

After a company performs any successful recruitment or signs any mutual contract with an employee, our proposed system adds a new block to the current Blockchain, by using **Algorithm 2**. Our proposed method used a hash-based data structure for storing data that is Merkle tree. In this system, data is hashed twice before storing [13-14]. All the data, securely stored in a Blockchain, such as a salary, title, promotion, training, leave, performance, transfer related contract (**BcHRMS**) can be used in the future for multi-faceted purposes including verification.

In short, we implemented proposed algorithms with Multichain open source software with the following characteristics. Identified block validator distributes the consensus. Although it's working procedure is similar to PBFT (Practical Byzantine Fault Tolerance). However, it does not validate by multiple validators but one block is validated by one validator. A round-robin consensus style where permitted miners generate blocks in round-robin style to generate and valid blockchain [32]. Section 3.2 elaborates the compliance achieving style of our proposed systems.

| Algorithm 2: BcHRMS (Blockchain based Human Resource Management System) |
|---|
| 1 **Input**: SL, Successful Applicants; |
| 2 C, Contract information |
| 3 **Output**: BC, updated Blockchain  // updated Blockchain |
| 4 **Begin** |
| 5 **for** each $SL_i$, i = 1:n do     //Successful recruitment |
| 6    Current Blockchain = Add new block(C, $SL_i$) |
| 7 **end** |
| 8 BC = Updated Blockchain |
| 9 **End** |

### 3.2 Proposed System Implementation Overview

For efficient development and use we divided our system into the following three (3) layers which are completely independent and scalable with current architecture:

**Decentralized/ distributed databases:** A decentralized or distributed database to store information. Typically one company will have one Blockchain node (i.e. one database) Examples: Oracle, MySQL, MerkleDB, MongoDB etc.

**Consensus Algorithms:** An algorithms to bring transparency and authenticity of the data. Examples include: Byzantine fault tolerance algorithm, the proof-of-stake algorithm (PoS) and the delegated proof-of-stake algorithm (DPoS).

**API and Applications:** For providing facilitates of human interactions through various applications (desktop, web and mobile). Application Programming Interface (API) shall play a vital role in this regard.

## 4  Implementation and Discussion

This section details the implementation of our proposed system and discusses the survey output results. A comparison of the proposed BcRMS and BcHRMS with existing recruitment and HRM systems has also been presented.

### 4.1 System Implementation

We implemented a prototype of our proposed systems (BcRMS and BcHRMS) using MultiChain 2.0 an open source Blockchain implementation platform [32-33]. The prototype was implemented on five Windows computer (windows 10, i7-6700 CPU @3.40Ghz Samsung Inc. South Korea) each representing a different participating entity such as recruiting company, previous/current employer, applicants, health sector and law agency. The justification for choosing MultiChain is it's globally used by many researchers [18]. Our proposed system architecture is a private Blockchain, where the network consensus is achievable with minimum difficulties and maintaining the trust between participants (Round Robin). Nodes can also be limited or defined. However, mining diversity is set to 0.75, following MultiChain implementation guideline. Although higher mining diversity can increase the security, it can also have negative effects if some miners become inactive for a prolonged period of time. A specific set of rules against forking were maintained methodically. An example of data storing (actual contract vs inside block data) is shown below (fig. 4):

Fig. 4: Contract information storing within a block (Blockchain data and Actual data)

### 4.2 Use Cases of Proposed Systems

**Basic information and identity management:** Personal information, last job, salary, health record, a criminal record can be verified from other institutions and organizations.

**Title and promotion management:** Selection of leader or position on the basis of past employers' reports and colleagues' opinions. A balanced and acceptable promotion and assignment of the title are possible by this system.

**Transfer and training management:** Training performance, results, ability, certificates, courses, etc. can be verified. Which is open to the participants for future verification.

**Incentive and salary management:** With a report from previous workplace and performance of current workplace salary and incentive are decided. This eliminates the chance of bribing or wrong information flow.

**Experience and achievement management:** Applicant's past experience, educational certificates, health information, salary structure, performance level i.e. every kind of achievements can be verified from actual authority.

### 4.3 Case Study

This section presents a survey result on 300 members of 12 medium to big organizations in South Korea and Bangladesh during the month of January and February, 2018 on proposed systems (**BcRMS**) and (**BcHRMS**). As shown in figure (fig. 5) around 60% of the participants agreed to proceed with our proposed approach whereas 20% consider the overall implementation to be expensive, 15% are of the opinion that they are not yet technically ready and another 10% consider more modification needed for adaptation.

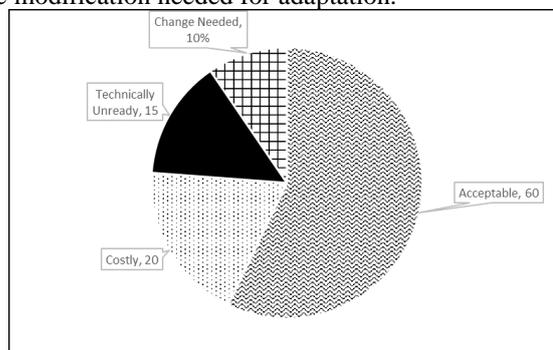

Fig. 5: Survey results of the proposed system

### 4.4 Comparative Analysis

To validate the proposed method we have compared them with existing recruitment and human resource management system in multiple aspects.

**Storage:** The Proposed system benefited from the advantages of the distributed database while a centralized database is used by most of the existing systems. Every participant (employees and employers) have full control on their information (assets) due to the separation of private and public key(s). No individual organization or employee can deny any particular verification and contract data. The system eliminates the possibility of Single Point of Failure (SPF) as it takes advantage of robust validation from several access points. Proposed system database information is almost impossible to edit or tamper. On the contrary, most of the current recruiting and human resource storage system has a single point of failure and operates in a less secure environment. Our system use Blockchain technology, which is a highly secure system known as the Trust Machine [34].

**Time:** On average, existent systems take several months to select an applicant for the company. If verification from several institutions is needed, the process further lingers. This phenomenon is also observed in completing other HRM related tasks. However, our proposed system can validate, verify and rank applicants' information in hours. Furthermore, once a contract agreement is finalized, a new block containing the information is added within few minutes. The time involved in getting the required consensuses from several nodes is basically the major share of the total processing time. This is far lower than the time taken by any other existing recruitment and HRM systems. Several studies have already stated the time needed for consensus (approval time) algorithm [13-14].

**Cost:** In our survey, 20 percent of surveyed participants consider the proposed system to be more expensive than existing systems. Our system needs a considerable amount of storage, has higher power consumption rate and that requires more investment during the initial setup process compared to the existing systems. However, in long-term, the return should be much higher than the investment. The major share of the expenditure of our proposed system is CAPEX (Capital Expenditure) whereas it is OPEX (Operational Expenditure) for the existing systems. Thus, our systems shall save cost in the long run. Furthermore, it shall also save any additional cost eventuated by wrong hiring using existing systems, as discussed in the introduction of this article.

**Quality:** We have already mentioned about the effect of wrong recruitment and biased HRM systems. In our proposed system, the quality of recruitment and HRM activities are ensured by machines. Our system can handle, store, validate and rank information with complete transparency and security. Since no fake and biased decisions are taken by our proposed system, there is no scope for quality degradation.

**Security:** Security of the existing system can easily be breached by hacking or system manager. Since Blockchain system are protected by double layered key and hashed encryption, information leaking and alteration of data is highly unlikely. Due to the distributed nature, a change in information can easily be tracked and original data can be retrieved without any major loss. As stated before, Blockchain thus provides an extra layer of security [13-14].

**Disintermediation:** Intermediate or third parties charges excessive cost for handling recruiting, verification, dispute settlement etc. Blockchain removes the need for middlemen and thus reduces expenditures of the companies. Moreover, the companies enjoy transparent and unbiased HRM system. Overall, our proposed system offers empowered users, higher quality and reliable recruitment and management, inter-process integrity, faster transaction and decision with lower cost. Below table. 1 shows a brief comparison between proposed systems and existing system used for recruitment and HRM [35].

| Factors | Existing System | Proposed System (BcRMS and BcHRMS) |
|---|---|---|
| Storage | Centralized | Decentralized |
| Encryption | Single layered | Double layered |
| Transparency | Internal | Customized |
| Security | Single point failure | No single point failure |
| Cost | High | Moderate to low |
| Time | High | Low |
| Overall Quality | Low | High |

Table 1: Comparison between current and proposed system

## 5 Conclusions

In this paper, to lower the risk of wrong recruitment and biased human resource management system, we proposed BcRMS and BcHRMS based on Blockchain technology. Blockchain is one of the key technologies for future smart industries of the fourth industrial revolution. Our Proposed models are capable of verifying and storing of recruitment and other HRM related information, offering a low-cost solution by removing the need for middlemen. Our research reveals that the proposed models can perform better compared to the existing HRM systems in terms of security, cost, time and quality of work. Furthermore, fake information and inappropriate promotion methods were identified by several surveys as two major problems in existent HRM systems, as discussed before. Using the proposed validation and verification algorithms, our proposed system successfully addressed those problems. Thus, the proposed system will have significant effects to build smart cities as well as towards smart industries in the era of existing Industry 4.0 and upcoming Industry 5.0. Our future research directions include developing a fully-fledged real-time application, followed by verification, dynamic update and system learning. This shall then enable the system to be commercially used in a broader scope with more accuracy.

**Acknowledgement:** Our deep gratitude to the anonymous reviewers for insightful comments. This work is supported by the Inje University research funds (Grant No.0001-2008-01000).

Md Mehedi Hassan Onik, Mahdi H. Miraz and Chul-Soo Kim, "A Recruitment and Human Resource Management Technique Using Blockchain Technology for Industry 4.0" in Proceeding of Smart Cities Symposium (SCS-2018), Manama, Bahrain, 2018, pp. 11-16.IET. "The link to the final version of the paper will be provided once it is included in IET." (Article in Press)